\documentclass[superscriptaddress,prx,twocolumn]{revtex4-2}
\usepackage{graphicx,amsmath,amssymb,physics,dsfont,subfigure}
\graphicspath{{}}
\usepackage[unicode=true,pdfusetitle,
 bookmarks=true,bookmarksnumbered=false,bookmarksopen=false,
 breaklinks=false,pdfborder={0 0 1},backref=false,colorlinks=false]
 {hyperref}
\usepackage[x11names]{xcolor}
\usepackage{orcidlink}

\makeatletter

\@ifundefined{textcolor}{}
{%
 \definecolor{BLACK}{gray}{0}
 \definecolor{WHITE}{gray}{1}
 \definecolor{RED}{rgb}{1,0,0}
 \definecolor{GREEN}{rgb}{0,1,0}
 \definecolor{BLUE}{rgb}{0,0,1}
 \definecolor{CYAN}{cmyk}{1,0,0,0}
 \definecolor{MAGENTA}{cmyk}{0,1,0,0}
 \definecolor{YELLOW}{cmyk}{0,0,1,0}
}

\hypersetup{breaklinks=true,colorlinks=true,citecolor=blue,linkcolor=blue,filecolor=blue,urlcolor=blue}
\IfFileExists{lmodern.sty}{\usepackage{lmodern}}{}

\makeatother

\begin{document}

\title{Large-scale quantum simulations of dissipative spin-1/2 Heisenberg chains}

\begin{abstract}

A quantum many-body system coupled to an environment relaxes to a
nonequilibrium steady state that can sustain order with no equilibrium
counterpart.  Computing such steady states is harder than
closed-system dynamics as the density matrix problem squares the
Hilbert-space dimension, and no free energy selects the steady state.
The dissipative spin-1/2 Heisenberg chain is a benchmark example for nonequilibrium steady state physics; various methods
have each calculated its phase diagram but do not agree, and a controlled
determination at large system size has remained out of reach.  Here we
simulate the Lindblad dynamics of chains of up to 50 sites on the
superconducting processor \texttt{ibm\_kingston}---100 simultaneously
active qubits at up to 1700 entangling-gate depths---realizing the dissipation via Stinespring dilation. 
The system's dissipative evolution is a self-correcting mechanism that effectively erases errors,
so hardware noise enters only as a weak competing dissipator.
We measure static
structure factors and resolve ferromagnetic, antiferromagnetic,
spin-density-wave, and paramagnetic steady states, mapping the phase diagram with 117 quantum hardware data points across the
$J_x$--$J_y$ plane. We uncover a rich non-equilibrium phase diagram of ordered phases with only remnants of the mean-field order, and where sharp transitions give way to the crossovers expected in one dimension. We also find the existence of an incipient (Trotter-induced) spin density wave phase, highlighting the potential of controlled Trotterization as a tool to engineer various magnetic phases in dissipative spin systems. 
Our quantum simulations largely settle the lingering uncertainty regarding the correct phase diagram of this benchmark system. Moreover, they show that quantum computers are now a feasible tool for addressing scientific questions involving dissipative quantum systems.

\end{abstract}

\author{Jo\~ao C. Getelina\orcidlink{0000-0002-1924-9813}}
\affiliation{Department of Physics and Astronomy, North Carolina State University, Raleigh, North Carolina 27695, USA}

\author{Andrew Cox\orcidlink{0009-0004-7062-6110}}
\affiliation{Department of Physics and Astronomy, North Carolina State University, Raleigh, North Carolina 27695, USA}

\author{Muhammad Asaduzzaman\orcidlink{0000-0001-7559-3873}}
\affiliation{Department of Physics and Astronomy, North Carolina State University, Raleigh, North Carolina 27695, USA}

\author{Omar Alsheikh\orcidlink{0009-0008-2012-5085}}
\affiliation{Department of Physics and Astronomy, North Carolina State University, Raleigh, North Carolina 27695, USA}

\author{Ryan S. Bennink\orcidlink{0000-0002-4810-9369}}
\affiliation{Computational Sciences and Engineering Division, Oak Ridge National Laboratory, Oak Ridge, Tennessee 37831, USA}

\author{James K. Freericks\orcidlink{0000-0002-6232-9165}}
\affiliation{Department of Physics, Georgetown University,
37th and O Sts. NW, Washington, DC 20057, USA}

\author{Alexander F. Kemper\orcidlink{0000-0002-5426-5181}}
\affiliation{Department of Physics and Astronomy, North Carolina State University, Raleigh, North Carolina 27695, USA}

\maketitle

When coherent many-body dynamics compete with a dissipation that violates detailed balance, a quantum
system relaxes not to a Gibbs state but to a nonequilibrium steady
state (NESS)~\cite{fazio2025many-body,sieberer2025driven-open}.  The dynamics and phase diagrams are governed by the Liouvillian
rather than by a free energy
~\cite{kessler2012central-spin,minganti2018spectral-theory}, 
and order without an equilibrium counterpart can be stabilized, including exotic
magnetism~\cite{lee2013unconventional-magnetism,jin2016cluster-meanfield}, time crystals~\cite{chinzei2020time-crystals}, limit cycles~\cite{owen2018limit-cycles}, synchronization~\cite{buca2022quantum-synchronization}, and dissipative
chaos~\cite{ferrari2025quantum-chaos}.  A
paradigmatic arena for these questions is the dissipative anisotropic spin-1/2 Heisenberg model with local spin decay, whose
mean-field phase diagram hosts ferromagnetic (FM), antiferromagnetic
(AFM), spin-density-wave (SDW), and paramagnetic (PM) steady states
meeting at Lifshitz points~\cite{lee2013unconventional-magnetism}.  A decade of scrutiny has
shown that correlations beyond mean field radically reshape this diagram: the
paramagnet becomes re-entrant and incommensurate tendencies
appear~\cite{jin2016cluster-meanfield,kshetrimayum2017tensor-network}. 
This model has become a 
benchmark for open-system
numerics~\cite{rota2017critical-behavior,nagy2019neural-network,hartmann2019many-body,vicentini2019steady-states,huybrechts2019cluster-methods,melo2025variational-perturbation,sander2025stochastic-simulation},
but as yet the phase diagram remains in question
because the numerics are limited;
the Lindblad equation~\cite{kossakowski1972non-hamiltonian,lindblad1976dynamical-semigroups,gorini1976completely-positive}
squares the Hilbert-space dimension, the NESS doesn't minimize a
free energy~\cite{weimer2021simulation-methods}, and quasi-exact methods reach only
small lattices or weak
entanglement~\cite{daley2014quantum-trajectories,verstraete2004matrix-product,zwolak2004superoperator-renormalization}.

Quantum computers are natural instruments for this problem, for two
reasons.  First, steady states
are attractors
of the dynamics, and can be reached from generic initial states without the
state-preparation overhead that burdens closed-system quantum
simulations~\cite{weimer2021simulation-methods,miessen2023quantum-dynamics}.  Second, and
counterintuitively, the physics is intrinsically noise correcting: a
dissipative system loses memory of its initial state, and of errors
accumulated along the way, at a rate set by the Liouvillian spectral
gap~\cite{minganti2018spectral-theory}, so hardware noise enters merely as an
additional dissipator competing with the engineered one---damping
observables, but not scrambling the structure from which phases are
identified~\cite{garciaperez2020ibmq-testbed,delre2020fermionic-reservoir,kamakari2022imaginary-time,cattaneo2023collective-effects,delre2024correlation-functions,rost2025error-mitigating}.
The dissipative character of the problem, the very source of its
classical hardness, is an asset on noisy hardware.

Beyond analog driven-dissipative
platforms~\cite{kasprzak2006exciton-polaritons,barreiro2011trapped-ions,carusotto2013quantum-fluids,helmrich2018nonequilibrium-phase,fitzpatrick2017circuit-qed,ma2019mott-insulator},
digital hardware experiments have so far either used dissipation
instrumentally---engineered reservoirs cooling spin systems toward the
low-energy states of target Hamiltonians~\cite{diehl2008cold-atoms,verstraete2009dissipation-engineering,mi2024engineered-dissipation,seki2026ground-state}---or probed
measurement-induced dynamics, as in the landmark adaptive-circuit
study of an absorbing-state transition by Chertkov \emph{et
al.}~\cite{chertkov2023non-equilibrium} and its
successors~\cite{pokharel2025adaptive-circuits,wu2026measurement-feedback}, where the steady state is a
trivial dark state and the critical physics lives in transients and
trajectory-to-trajectory fluctuations.  Deterministic Lindblad
evolution toward a nontrivial ensemble-averaged NESS, scanned across a
many-body phase diagram, has not been realized on quantum hardware at
any scale.

\begin{figure*}[t]
    \centering
    \includegraphics[width=\linewidth]{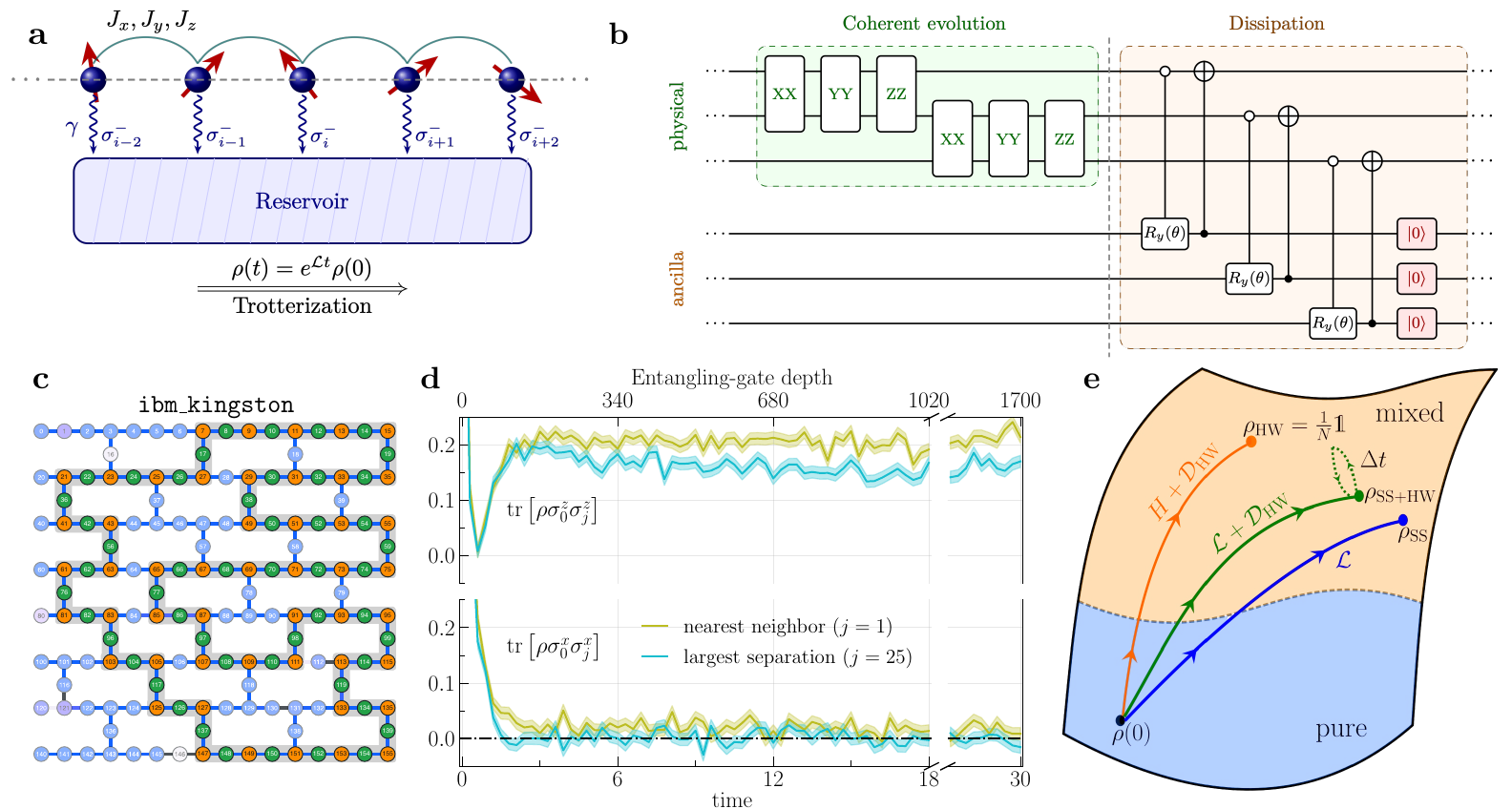}
    \caption{\textbf{Stable non-equilibrium steady states of the dissipative Heisenberg model}
    (a) 1D dissipative spin-1/2 Heisenberg chain. The system sites interact with their nearest neighbors 
    and with a reservoir that flips the spin down
    at a dissipation rate $\gamma$. 
    (b) Quantum circuit representation of a single time step in the Liouvillian dynamics of an open-boundary chain of size $L=3$, consisting of two parts:
    a coherent evolution under the closed system Hamiltonian, and a (non-unitary) dissipative part representing the interaction with the reservoir, which is implemented by coupling to an ancilla that is reset each time step.
    (c) Qubit layout used in our simulations in the 156-qubit superconducting quantum device \texttt{ibm\_kingston}. Here show a closed chain with $L=50$ sites (100 qubits including ancillae). Green (orange) dots represent the physical sites (ancilla) of active qubits. 
    (d) Stable non-equilibrium steady state from \texttt{ibm\_kingston} for the nearest neighbors and the two furthest apart sites for $(J_x,J_y)=(3,2)$, and with $J_z=\gamma=1$. The shaded area around each curve corresponds to the standard error of the mean.
    (e) Conceptual underlying reason for the stability. The hardware noise shifts the steady state from the model $\rho_\textrm{SS}$ to a modified $\rho_\textrm{SS+HW}$; as long as the model evolution is sufficient to balance the hardware noise, the $\rho_\textrm{SS+HW}$ will retain the physics of $\rho_\textrm{SS}$.
    }
    \label{fig:summary}
\end{figure*}

Here we realize precisely this.  We simulate dissipative spin-1/2 Heisenberg
chains of up to $L = 50$ sites ($100$ qubits) on the superconducting processor
\texttt{ibm\_kingston} in circuits of up to 1700 entangling-gate depth. We employ a Trotterization of the Liouvillian to first order
and realize the dissipator via the Stinespring dilation, with ancillary qubits that are reset after every time step~\cite{rost2025error-mitigating,delre2024correlation-functions,vu2025oscillator-qubit,delmonico2026teaching}. This approach has the advantage of being simple to implement compared to competing unitary embeddings~\cite{childs2017sparse-markovian,cleve2017lindblad-evolution,miessen2023quantum-dynamics,ding2024hamiltonian-simulations,borras2025lindblad-simulation}.
The measured correlation matrices yield the static structure factor $S(q)$, whose
peak structure identifies FM, AFM, SDW, and PM behavior. From 117
hardware data points across the $J_x$--$J_y$ plane we sketch the
steady-state phase diagram, finding qualitative agreement with
mean-field theory but only when deep in the ordered phases~\cite{lee2013unconventional-magnetism}.
Since the studied model is one dimensional, where no spontaneous symmetry-breaking phase transition is expected,
we find crossovers between finite-size remnants of the mean-field orders~\cite{jin2016cluster-meanfield}.\\

\textit{Dissipative spin-1/2 Heisenberg model.}
The dissipative anisotropic spin-1/2 Heisenberg Hamiltonian in 1D is given by
\begin{equation}\label{eq:hamiltonian}
    H = \frac{1}{2}\sum_{\langle i,j\rangle}
    \left[ J_x\sigma^x_i\sigma^x_j+J_y\sigma^y_i\sigma^y_j+J_z\sigma^z_i\sigma^z_j
    \right],
\end{equation}
where $\sigma^\alpha_i$ are Pauli operators, $J_\alpha$ are the coupling amplitudes. By coupling this system to a memoryless (i.e., Markovian) environment through local spontaneous spin decay, the corresponding Lindblad master equation that governs its dynamics becomes~\cite{kossakowski1972non-hamiltonian,lindblad1976dynamical-semigroups,gorini1976completely-positive},
\begin{align}
\label{eq:gksl}
    \dot{\rho}(t) =& -i\left[H,\rho(t)\right] + \mathcal{D}\nonumber\\
    \mathcal{D} =& \gamma\sum_m \left( L_m\rho(t)L^\dagger_m-\frac{1}{2}\left\{L^\dagger_m L_m,\rho(t)\right\} \right),
\end{align}
with jump operators $L_m=\sigma^-_m=(\sigma^x_m-i\sigma^y_m)/2$ acting on every site $m$ at rate $\gamma$ [Fig.~\hyperref[fig:summary]{\ref{fig:summary}(a)}].
For this model, single-site mean-field theory predicts that the steady state hosts FM, AFM, SDW, and PM phases meeting at Lifshitz points across the $J_x$--$J_y$ plane at fixed $J_z=\gamma=1$~\cite{lee2013unconventional-magnetism}. This phase diagram, and its fate beyond mean field~\cite{jin2016cluster-meanfield}, is the target of our quantum hardware simulations.\\

\textit{Quantum hardware simulations.}
\label{sec:results}
We performed all the quantum simulations on the 156-qubit superconducting device \texttt{ibm\_kingston}, which is depicted in Fig.~\ref{fig:summary}(c), using up to 100 qubits.
We implemented an alternating pattern between physical and ancilla qubits (green and orange circles in Fig.~\hyperref[fig:summary]{\ref{fig:summary}(c)}, respectively), yielding an entangling-gate circuit depth of 17 per Trotter step. We choose the initial state according to the target observable, assigning a fully-polarized up state in the same basis as the measurement; i.e., the initial state is either $|0\rangle^{\otimes L}\otimes|0\rangle^{\otimes L}$ or $|0\rangle^{\otimes L}\otimes|+\rangle^{\otimes L}$ (with $\sigma^x_i|\pm\rangle=\pm|\pm\rangle$), when measuring ZZ- and XX-correlators, respectively, with the leftmost $L$ qubits corresponding to the ancillae.

To estimate the steady state, we evolve the initial state to a sufficiently long time, up to $t=30$, with a time step $\Delta t=0.3$, resulting in an entangling-gate depth of 1700. Fig.~\hyperref[fig:summary]{\ref{fig:summary}(d)} shows the time trace of the ZZ- and XX-correlation between the two closest and the two furthest apart sites for a periodic chain of size $L=50$ [Fig.~\hyperref[fig:summary]{\ref{fig:summary}(c)}]. Notice that the ZZ-correlation for both distances settles at a non-zero value, which does not degrade as the circuit deepens, thus corroborating our assertion about the resilience of open-system dynamics against hardware noise. On the other hand, the XX-correlation, which is the quantity used for identifying magnetic phases, appears to reach a non-zero steady-state value only for the nearest-neighboring pair.

A time step of $\Delta t=0.3$ is large enough to raise Trotter-error concerns, yet under hardware noise we observe the opposite of the noiseless expectation: a larger time step brings the system \emph{closer} to the target NESS. The reason for this is a competition between dissipators [Fig.~\hyperref[fig:summary]{\ref{fig:summary}(e)}]: noise converts the ideal evolution $\mathcal{L}$ into $\mathcal{L}+\mathcal{D}_\textrm{HW}$, where $\mathcal{D}_\textrm{HW}$ is the unknown hardware dissipator,
and the steady state that dominates at long times is the one whose Liouvillian gap is larger. Because the engineered dissipation enters each circuit through the rotation angle $\theta(\gamma\Delta t)$, while the hardware noise per step is fixed, increasing $\Delta t$ strengthens the target dissipator relative to $\mathcal{D}_\textrm{HW}$, at the price of additional Trotter error (see App.~\ref{app:timestep} and Ref.~\cite{prampain2026trotter-universality}).

\begin{figure}[t]
    \includegraphics[width=0.49\textwidth]{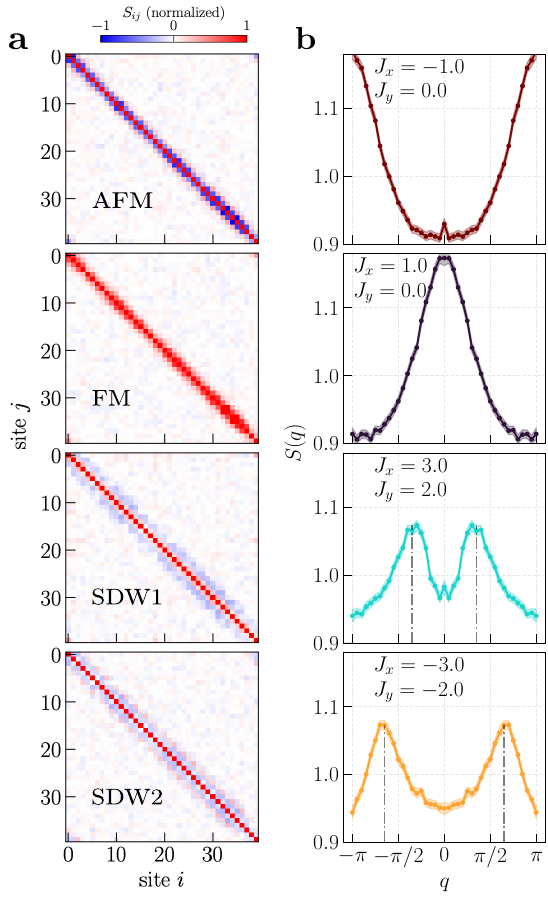}
    \caption{\textbf{Distinguishing steady state phases of the dissipative Heisenberg model.} Real space (a) and momentum space (b) structure factors for a 40 site (80 qubit) chain, averaged over the last 10 time steps. Vertical lines in the SDW panels in (b) indicate the expected ordering vector for tensor network calculations for 40 sites. Shaded areas around the curves correspond to the 95\% confidence interval of 1000 bootstrap samples, resampled from the last 10 time steps.
    }
    \label{fig:ness_phases}
\end{figure}

\begin{figure}[t]
    \centering
    \includegraphics[width=0.49\textwidth]{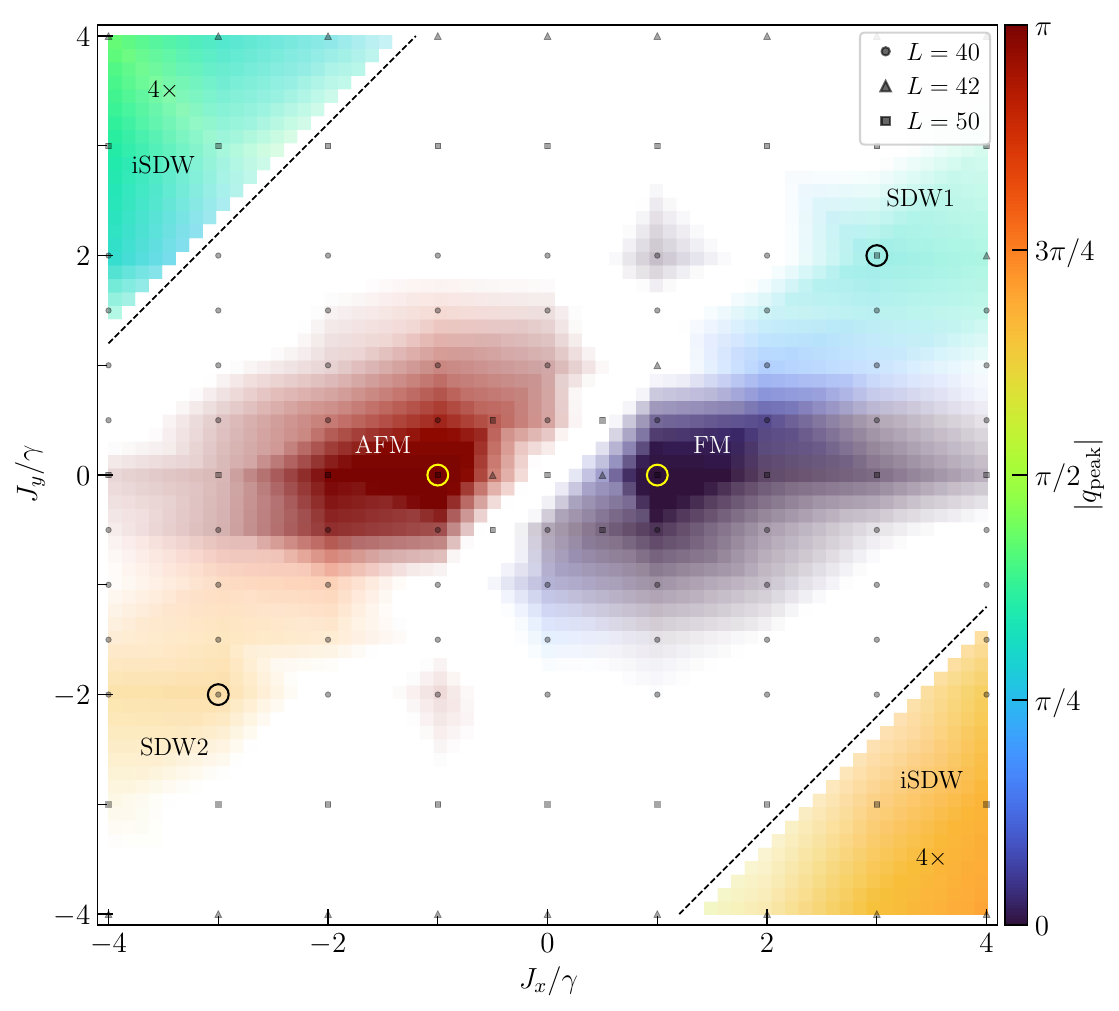}
    \caption{\textbf{Phase diagram of the dissipative spin-1/2 Heisenberg model.} Peak position in $S(q)$ from quantum hardware simulation, opacity is set by the height of the peak. Data is interpolated before plotting. Corners show an incipient SDW phase, which has been scaled $4\times$ for visibility. All correlator data obtained from the hardware used to create this phase diagram are presented in the App.~\ref{app:all-hw-results}.
    \label{fig:phase-diagram}
    }
\end{figure}

\textit{Phase diagram.}
To find the phase diagram across the $J_x$--$J_y$ plane (fixed $J_z=\gamma=1$), we measure the static XX correlation function, which is given by
\begin{align}
S_{ij}=\mathrm{Tr} \left[\rho_{\mathrm{SS}}\sigma^x_i\sigma^x_j\right],\label{eq:xx-corr}
\end{align}
where $\rho_\textrm{SS}$ is the approximate non-equilibrium steady state, which is estimated by letting the system evolve for a relatively long time.
In Fig.~\ref{fig:ness_phases}(a) we show the full set of static correlation functions at the end of the evolution for all pairs of sites $i,j$ on a 40-site periodic chain (80 qubits) and for 4 separate points in the $(J_x,J_y)$ phase diagram. Here we have averaged over the last ten circuits of the evolution ($t \in \left[27.3, 30\right]$) to mitigate the effects of shot noise. Notice that the magnetic phases are already distinguishable at this level: the ferromagnetic (FM) phase looks notably different from the anti-ferromagnet (AF) and spin density wave (SDW). This is also reflected in the momentum space static structure factor (Fig.~\hyperref[fig:ness_phases]{\ref{fig:ness_phases}(b)}), which is defined as
\begin{align}
  S(q)=\frac{1}{L}\sum_{i,j=1}^L e^{iq(r_i-r_j)}S_{ij},\label{eq:sq}
\end{align}
whose peak position distinguishes the candidate orders: $q=0$ (FM), $q=\pi$ (AFM), incommensurate $q$ or spin density wave (SDW), or no peak above baseline set by the $r_i=r_j$ terms (PM).
According to the mean-field predictions of Ref.~\cite{lee2013unconventional-magnetism}, the four selected coupling pairs of Fig.~\hyperref[fig:ness_phases]{\ref{fig:ness_phases}} should correspond to AFM, FM and SDW phases. Indeed, this is what we observe in Fig.~\hyperref[fig:ness_phases]{\ref{fig:ness_phases}(b)}, with two curves peaking at commensurate momenta and the other two peaking at an incommensurate momenta. These results also agree with small-scale ($L=12$) exact numerical calculations (see App.~\ref{app:exact-sims-L12}), and same scale ($L=40$) tensor network simulations (see App.~\ref{app:tn-sims-L40}), from which we show the resulting peak locations (dot-dashed lines) for the SDW phases.

We now move beyond these representative points and scan the phase diagram. To distinguish between the different magnetic phases, we keep track of the height and position of the peak in $S(q)$ calculated from the quantum hardware data.
Fig.~\ref{fig:phase-diagram} shows the resulting phase diagram, interpolated from 117 hardware data points for three different system sizes, namely $L=40,\,42,$ and $50$
One can find details about the employed interpolation scheme in App.~\ref{app:interpolation}. 
We clearly distinguish the ferromagnetic (FM, $q=0$), anti-ferromagnetic (AFM, $q=\pi$) and spin density wave (SDW, incommensurate $q$) phases, as well as the paramagnetic regions where there is a lack of order in the XX correlators.
The FM and AFM regions are primarily near the $J_y=0$ line, and are appropriately symmetric under $(J_x,J_y) \rightarrow (-J_x,-J_y)$ due to the inherent symmetry in the model.
The SDW phases are clearly visible, with two identifiable regions of longer (light green, upper-right corner) and shorter (light orange, lower-left corner) spatial modulations. Coincidentally, these areas are in good agreement with the mean-field predictions of Ref.~\cite{lee2013unconventional-magnetism}.
The most striking difference from the mean-field predictions~\cite{lee2013unconventional-magnetism} is the absence of a large PM area near the center of the diagram; rather, this is limited to the region right around $J_x=J_y$ where the fixed point is the dark state 
$|1\rangle^{\otimes L}$.
Moreover, there is a re-entrant paramagnetic phase---appearing and disappearing across the phase diagram---at large couplings that is driven by a much different mechanism: as the interactions grow stronger, the state becomes increasingly mixed, eventually losing all transverse order.

\textit{Incipient SDW ordering.}
We observe a signal that indicates SDW ordering in the upper left and lower right corners of the phase diagram, which is not observed in either the mean field phase diagram or the $L=12$ numerical results. However, rather than this being a feature that only appears at large system sizes, we attribute this to an effective Hamiltonian that is modified from the dissipative Heisenberg model of Eq.~\ref{eq:gksl} due to Trotter error. When $\Delta t$ is large, commutator terms arise from the Trotter decomposition that behave like additional effective terms in the Hamiltonian. In certain instances this has proven to be a shortcut to adding desired terms that are hard to implement~\cite{ali2025robust}; here, however, it modifies the Hamiltonian in the far corners of the phase diagram (where $|J|$ is large and thus Trotter error is worse), and we start to see the phases that arise from it. For the dissipative Heisenberg model, the additional terms that appear centered on the $m^\mathrm{th}$ site are the chiral terms of the form
    $Z_{m-1} Y_m X_{m+1} - X_{m-1} Y_m Z_{m+1}$
and its cyclic permutations.
Indeed, these enhance an incipient SDW phase, as confirmed by small exact diagonalization calculations including these terms (see App.~\ref{app:incipient-sdw}).  While this is not strictly part of the dissipative Heisenberg model, it does demonstrate that the remarkable stability and conceptual framework of a modified $\rho_{\mathrm{SS}}$
carries over to extensions away from this model.

\begin{figure}[t]
    \centering
    \includegraphics[width=0.45\textwidth]{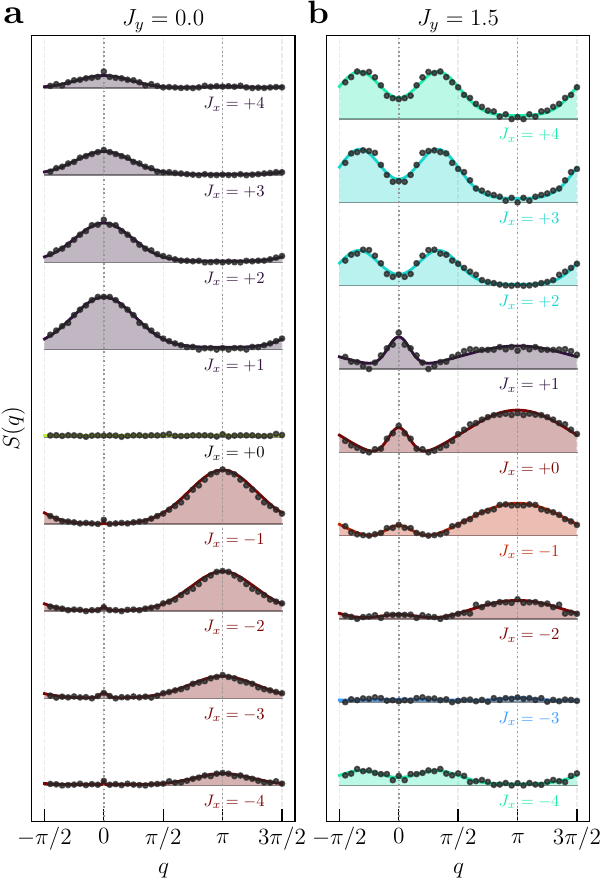}
    \caption{\textbf{Observing the crossover between magnetic phases.} Static structure factor $S(q)$ for several values of coupling amplitude $J_x$ and fixed (a) $J_y=0.0$ and (b) $J_y=1.5$. Symbols correspond to the quantum hardware data points, whereas lines and shaded area correspond to curve fits of sum of Gaussian distributions. System size is $L=40$ (80 qubits) and the dissipative rate is $\gamma=J_z=1$.
    \label{fig:qcuts}
    }
\end{figure}

Finally, we also consider two horizontal cuts in the phase diagram, and investigate how the corresponding static structure factor changes as a function of $J_x$. Figs.~\hyperref[fig:qcuts]{\ref{fig:qcuts}(a)} and \hyperref[fig:qcuts]{(b)} correspond to horizontal cuts at $J_y=0$ and $J_y=1.5$, respectively. We compare the quantum hardware data points (gray symbols) of the static structure factor with respect to Gaussian fits (lines and shaded areas) as we decrease the value of $J_x$ (top to bottom). For $J_y=0$ [Fig.~\hyperref[fig:qcuts]{\ref{fig:qcuts}(a)}], notice how the system gradually changes from a ferromagnet to an antiferromagnet, separated by a paramagnetic phase at $J_x=0$. On the other hand, for $J_y=1.5$ [Fig.~\hyperref[fig:qcuts]{\ref{fig:qcuts}(b)}], the system starts at an incommensurate $q$ phase until it shifts to a state with coexistence of FM and AFM phases, with the former losing its dominance to the latter as we decrease $J_x$. Eventually, however, both the FM and AFM peaks become negligible, and the system enters a PM phase at $J_x=-3$, which is then replaced by an incipient SDW phase at $J_x=-4$, arising due to higher-order corrections in the Trotter approximation (see App.~\ref{app:incipient-sdw}).

\textit{Discussion and outlook.}
After a decade in which each new classical method redrew the phase
diagram of the dissipative Heisenberg chain, our results in Fig.~\ref{fig:phase-diagram}
supply a measure of ground truth. At $L=40$--$50$
the steady state organizes into finite-size remnants of the mean-field
orders separated by crossovers, with the paramagnet confined to the
vicinity of the dark-state line $J_x=J_y$ rather than the broad
central region predicted by mean-field
theory~\cite{lee2013unconventional-magnetism}.

The quantum hardware result rests on the stability mechanism of
Fig.~\hyperref[fig:summary]{\ref{fig:summary}(e)}: the engineered
dissipation continually erases accumulated errors, so hardware noise
enters as a weak additional dissipator and the measured steady state
$\rho_{\mathrm{SS+HW}}$ retains the structure of the target
$\rho_{\mathrm{SS}}$.  The same mechanism sets the method's
boundaries.  The competition between the two dissipators is governed
by the ratio of the dissipation, Trotter step, and hardware error [$\gamma\Delta t/\varepsilon$, Eq.~\eqref{eq:ratio}], and
it is lost where large coupling amplitudes amplify the Trotter error,
as in the far corners of the diagram.  Moreover, because our
diagnosis rests on the static structure factor, we can locate the
crossovers between magnetic behaviors but cannot certify a phase
boundary; in one dimension none is expected, but sharpening this
distinction will require quantitative error mitigation beyond the
suppression techniques employed here.

The incipient SDW phases in the upper-left and lower-right corners
carry some implication beyond the fact that they are observed there.
The phases originate from chiral three-site
terms generated by the Trotter decomposition, effectively changing the Liouvillian being simulated. Typically such terms produce phases absent from the target model and are considered to be a problem with the simulation.
However, applied deliberately, they realize interactions that are
otherwise costly to implement~\cite{ali2025robust} and can even tune
the universality class of a dissipative
process~\cite{prampain2026trotter-universality}.  Since the stability
mechanism is agnostic to which Liouvillian is simulated,
Trotter-engineered extensions of this kind inherit it at no
additional cost.

Our data also set the terms for comparison to state of the art classical methods.  The 117 measured
points, and the underlying $S(q)$ curves at $L=40$--$50$, constitute
a quantitative target for the classical contenders: the bond
dimension at which the tensor-jump
method~\cite{sander2025stochastic-simulation}---which reached 1000
spins at bond dimension four---reproduces them is now a well-posed
question, as is whether
variational-perturbative~\cite{melo2025variational-perturbation} and
neural-network~\cite{nagy2019neural-network,hartmann2019many-body,vicentini2019steady-states}
ans\"atze agree.  The hardware systematics (noise and Trotter error)
are independent of the classical ones (bond dimension and ansatz
bias), so agreement between the two carries information about the
model rather than about either method.  This cross-validation,
more than any single calculation, will settle the phase diagram.

Finally, our results indicate that quantum hardware is now a feasible tool for addressing questions in dissipative and driven-dissipative systems.  The issues with decoherence and scaling that plague closed (unitary) dynamics are ameliorated by the inherent stability of a dissipative steady state, thus realizing the idea of accepting the dissipation rather than fighting it~\cite{verstraete2009dissipation-engineering,cubitt2023dissipative}.
The questions are now how such an approach can be applied to other problems of interest. Moving beyond one dimension---where true dissipative transitions are
expected~\cite{jin2016cluster-meanfield,rota2017critical-behavior,kshetrimayum2017tensor-network}
and classical methods are most strained--- is trivial within this framework, 
as the methods directly carry over. Driven models, such as those that exhibit limit cycles or time crystals, offer additional areas for exploration with quantum hardware simulation.

\section*{Acknowledgments}

We acknowledge useful discussions with Raghav Jha, Berkley Delmonico, and Yan Wang. This work was supported by the U.S. Department of Energy, Office of Science, Office of Advanced Scientific Computing Research under Award Number DE-SC0025623.  Exact numerical calculations were supported by the U.S. Department of Energy, Office of Science, Office of Advanced Scientific Computing Research under Award Number DE-SC0025430.

\section*{Methods}
\subsection*{Quantum hardware implementation}
To realize Eq.~\eqref{eq:gksl} on hardware, we write the time-independent Liouvillian as $\mathcal{L}=\mathcal{H}+\mathcal{D}$, where $\mathcal{H}[\rho]$ and $\mathcal{D}[\rho]$ are the right-hand side of Eq.~\eqref{eq:gksl} with $\gamma=0$ and $H=0$, respectively, and apply a first-order Trotter approximation to $\rho(t)=e^{\mathcal{L}t}\rho(0)$,
\begin{equation}
    \rho(t)\approx \left[e^{\mathcal{D}\Delta t} e^{\mathcal{H}\Delta t}\right]^{N}\rho(0),\label{eq:trotter}
\end{equation}
with time step $\Delta t=t/N$. The coherent factor is a standard Hamiltonian-simulation circuit, whereas the dissipative factor is realized exactly by a Stinespring dilation~\cite{vu2025oscillator-qubit}: each system qubit is coupled to a dedicated ancilla through a controlled-$R_y(\theta)$ rotation and a CNOT, with $\theta=2\sin^{-1}\left(\sqrt{1-e^{-\gamma \Delta t}}\right)$, and the ancillae are reset after every Trotter step [Fig.~\ref{fig:summary}(b)]. This is the simplest of the unitary embeddings of Lindblad dynamics: it doubles the qubit count, but adds only two entangling gates per site per step, and with physical and ancilla qubits alternating along the device the entangling-gate depth is 17 per Trotter step [Fig.~\ref{fig:summary}(c)].

All quantum hardware simulations use two error suppression techniques provided by the Qiskit Runtime service. First, Pauli twirling (randomized compiling)~\cite{wallman2016noise} is applied to every two-qubit entangling gate: each gate is sandwiched between randomly chosen single-qubit Pauli operators whose net action is the identity in the absence of noise, so that averaging over multiple random instances converts coherent gate errors, which accumulate quadratically with circuit depth, into a stochastic Pauli noise channel that accumulates only linearly~\cite{hashim2021randomized}. 
Second, dynamical decoupling is enabled with the XY4 pulse sequence~\cite{maudsley1986modified,viola1999dynamical}, which inserts a repeating X-Y-X-Y gate pattern on idling qubits during scheduling gaps in the circuit. The XY4 sequence suppresses low-frequency dephasing and residual ZZ crosstalk while being self-correcting against systematic over- or under-rotations in the decoupling pulses themselves~\cite{ezzell2023dynamical}. Neither technique adds ancilla qubits or additional two-qubit gates, so the overhead is limited to single-qubit operations that are fast compared to the dominant two-qubit gate errors. No error \emph{mitigation} techniques (such as zero-noise extrapolation or probabilistic error cancellation) are applied; the reported expectation values are raw hardware measurements with suppression only.

\bibliographystyle{apsrev4-2}
\bibliography{kemperlab}

\clearpage
\onecolumngrid
\appendix

\renewcommand\thefigure{\thesection\arabic{figure}}  
\setcounter{figure}{0}

\section{Effects of time-step size in the Liouvillian dynamics}
\label{app:timestep}
\setcounter{figure}{0}

In the absence of noise, the Trotter step $\Delta t$ controls only the
accuracy of the first-order product-formula approximation to
$e^{\mathcal{L}t}$, and smaller is always better.  Under hardware
noise this logic inverts.  Each Trotter step is realized by a circuit
layer of fixed entangling-gate depth and therefore deposits a fixed
amount of unwanted dissipation per step, which we characterize by a
strength $\varepsilon$.  Advancing the simulation to time $t$ requires
$N = t/\Delta t$ layers, so in simulated-time units the noise acts at
an effective rate
\begin{equation}
  \gamma_{\rm HW} \sim \frac{\varepsilon}{\Delta t},
\end{equation}
while the engineered dissipation acts at the fixed rate $\gamma$.
Which of the two competing steady states dominates at accessible times
is controlled by the dimensionless ratio
\begin{equation}
  \frac{\gamma}{\gamma_{\rm HW}} \sim
  \frac{\gamma\,\Delta t}{\varepsilon}.
  \label{eq:ratio}
\end{equation}
Increasing $\Delta t$ therefore strengthens the target Liouvillian
relative to the hardware noise floor, at the cost of a first-order
coherent
Trotter error that grows with $\Delta t$ and with the coupling
amplitudes $J_\alpha$.  For too small $\Delta t$ the measured state
approaches the noise fixed point; for too large $\Delta t$ the Trotter
error corrupts the coherent evolution, which is the origin of the
anomalous points at $|J_\alpha| \gtrsim 4$ discussed in Appendix~\ref{app:incipient-sdw}.  We note that the discretization is not merely a
source of numerical error: for the quantum contact process, the choice
of Trotterization can even tune the universality class of the
resulting dynamics~\cite{prampain2026trotter-universality}.  The time step must therefore
be regarded as a physical parameter of the simulated process, to be
chosen---as here---by balancing competing physical effects rather than
minimized indiscriminately.

We can illustrate this with a simple example.
Consider a
single spin with engineered decay toward $|\downarrow\rangle$ 
(jump operator
$\sigma^-$, rate $\gamma$, no Hamiltonian), and model the per-layer
hardware noise as a depolarizing channel of strength $\varepsilon$.
Per layer, the engineered channel acts with probability
$p = 1 - e^{-\gamma \Delta t} \approx \gamma\Delta t$, pulling the
state toward the target NESS 
$|\downarrow\rangle\!\langle \downarrow|$, while the noise
pulls the state toward the maximally mixed state $\openone/2$.  The
fixed point of the composite map is
\begin{equation}
  \tilde{\rho}_{\rm ss}
  = C\, |\downarrow\rangle\!\langle \downarrow| + (1 - C)\,\frac{\openone}{2},
\end{equation}
\begin{figure}[b]
    \includegraphics[scale=0.5]{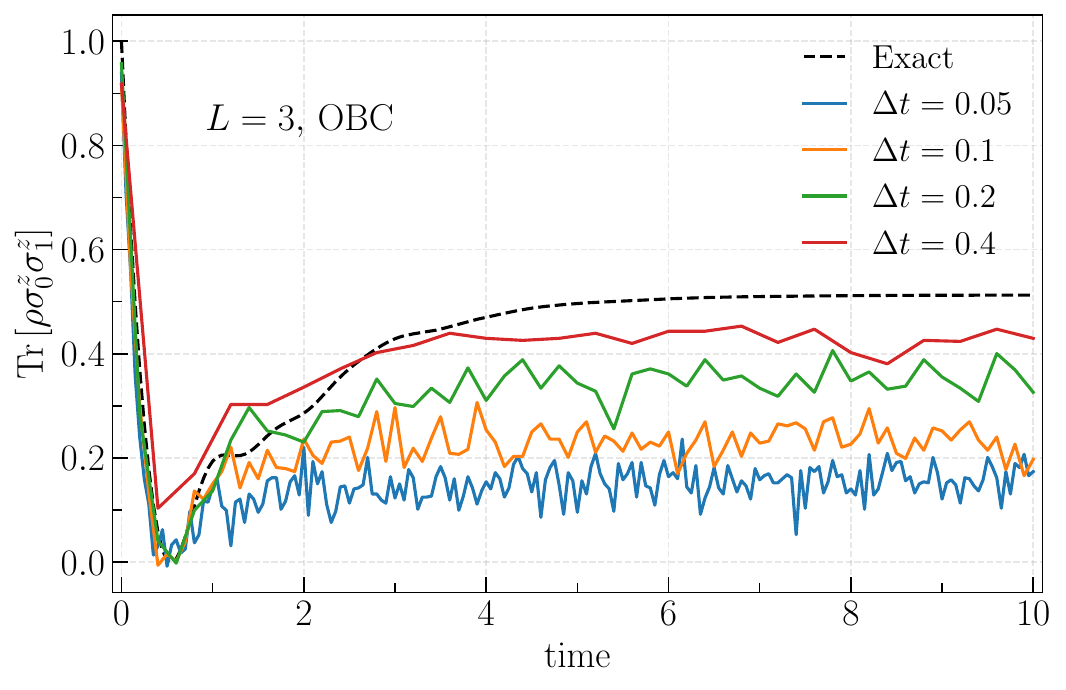}
    \caption{
    Time evolution of the ZZ correlation function between nearest neighbors in a dissipative Heisenberg chain of size $L=3$, with coupling amplitudes $(J_x,J_y)=(4,2)$ and dissipative rate $\gamma=J_z=1$. The dashed black line corresponds to exact simulations, whereas the remaining curves correspond to quantum circuit simulations with variable time step on a fake backend emulating the error rates of the $\texttt{ibm\_fez}$ device.
    }
    \label{fig:compare-dt}
\end{figure}
with contrast
\begin{equation}
  C = \frac{p}{p + \varepsilon(1-p)}
    \approx \frac{\gamma \Delta t}{\gamma \Delta t + \varepsilon},
\end{equation}
i.e., the target NESS survives with a contrast that increases
monotonically with $\Delta t$.  This is precisely the behavior we
observe in the interacting model: hardware noise suppresses the
amplitude of the correlations [Fig.~\hyperref[fig:summary]{\ref{fig:summary}(e)}] without
destroying the structure from which the magnetic phase is identified,
and larger time steps improve the contrast.

To reinforce this idea of the inverted logic regarding time-step size for noisy simulations, we provide some numerical results for the dissipative Heisenberg chain using a noise model that emulates the error rates of real quantum hardware. Fig.~\ref{fig:compare-dt} shows the time evolution of the ZZ correlation between nearest-neighboring sites in an open boundary chain of size $L=3$. We compare the exact simulation results (dashed black line) with quantum circuit simulations on the fake backend that emulates the device $\texttt{ibm\_fez}$, for which we consider four different values of the time step $\Delta t$. Notice that as we make the time step larger, the curves approach the exact results, which is the opposite behavior as expected in the noiseless case $\gamma_\textrm{HW}=0$. Moreover, since for these simulations $H\neq0$, the role of the engineered dissipation $\gamma$ in the previous example is now played by the Liouvillian gap.

\section{Compilation of quantum hardware results}\label{app:all-hw-results}
\setcounter{figure}{0}

This appendix presents the complete set of raw measurement data obtained from the dissipative Heisenberg simulations on quantum hardware.
\begin{figure}
    \includegraphics[width=0.9\textwidth]{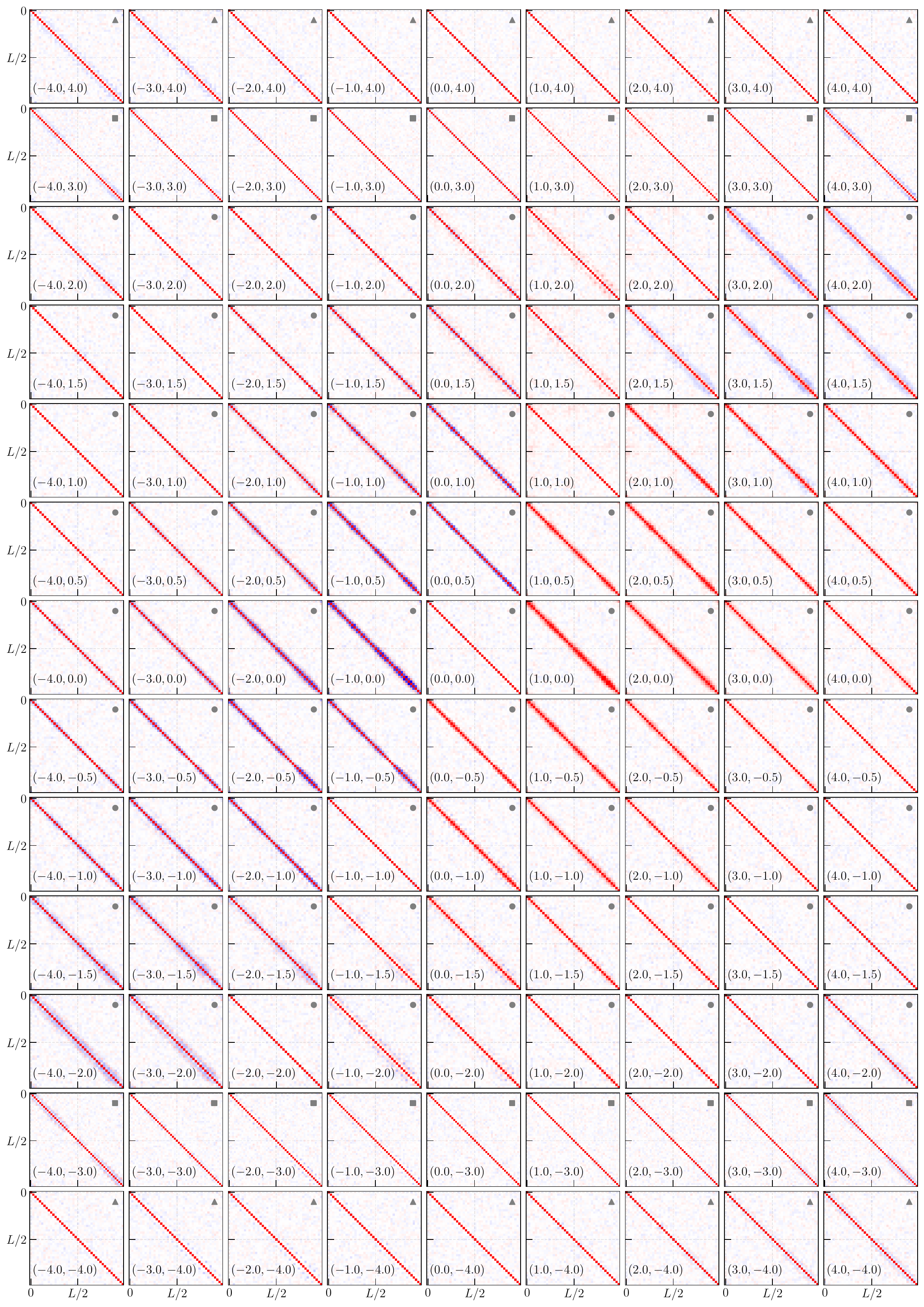}
    \caption{
    Full grid of quantum hardware data for the (normalized) steady-state correlation matrix $S_{ij}=\mathrm{Tr} \left[\rho_{\mathrm{SS}}\sigma^x_i\sigma^x_j\right]$
    for several coupling amplitude pairs $(J_x,J_y)$ and periodic chain length identified by the upper-right symbols: $\text{\Large$\bullet$}\rightarrow L=40$, $\blacktriangle \rightarrow L=42$, and $\blacksquare \rightarrow L=50$. Data corresponds to the average over the last 10 time steps, with $t \in [27.3,30]$ and dissipative rate $\gamma = J_z = 1.0$.
    }
    \label{fig:cm-grid}
\end{figure}
\begin{figure}
    \includegraphics[width=0.9\textwidth]{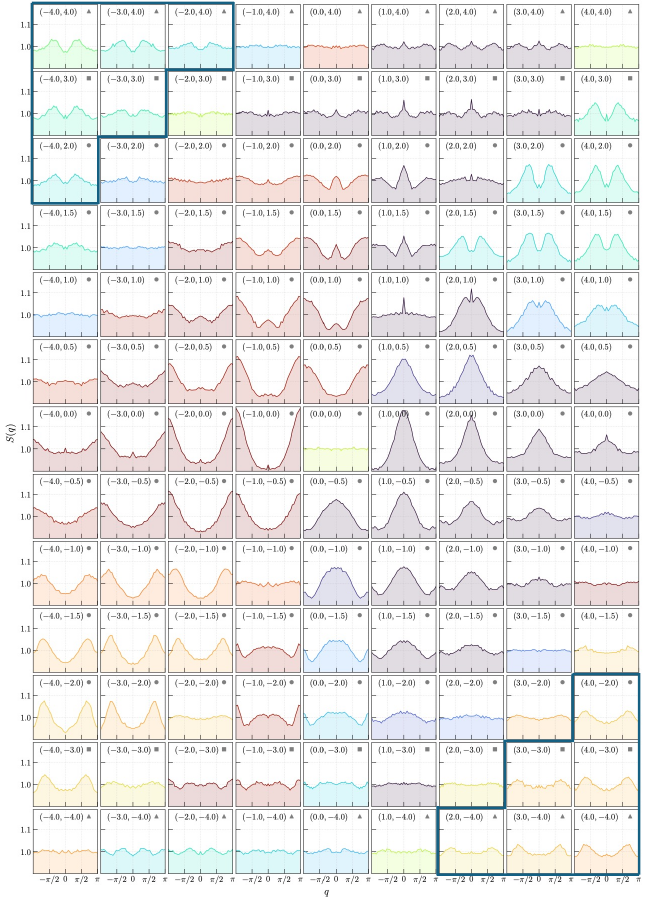}
    \caption{
    Same as Fig.~\ref{fig:cm-grid}, but with the correlation measurement computed in momentum space, i.e., $S(q)=\frac{1}{L}\sum_{i,j=1}^L e^{iq(r_i-r_j)}S_{ij}$. Color of the shaded areas reflect the magnetic phase the system is found, following the same color scheme as the phase diagram reported in the main text (Fig.~\ref{fig:phase-diagram}). System size is indicated by the marker shape.
    }
    \label{fig:sq-grid}
\end{figure}
In Figs.~\ref{fig:cm-grid} and~\ref{fig:sq-grid} we present our complete quantum hardware dataset as correlation matrices ($S_{ij}$, Eq.~\ref{eq:xx-corr}) and static structure factors ($S(q)$, Eq.~\ref{eq:sq}), respectively. These data span 117 points across the parameter space, from which the phase diagram reported in the main text (Fig.~\ref{fig:phase-diagram}) is drawn.
Overall, the hardware reproduces the qualitative structure expected from mean-field predictions~\cite{lee2013unconventional-magnetism}: ferromagnetic (FM) and antiferromagnetic (AFM) regions near the line $J_y=0$, for $J_x>0$ and $J_x<0$, respectively, and two spin-density-wave (SDW) wedges with distinct spatial modulation from each other. However, as discussed in the main text, there are some notable changes in the phase diagram compared to the mean-field predictions. Here we focus on the two most striking ones: the existence of a re-entrant paramagnetic (PM) phase and the incipient SDW phases.

One can observe the re-emergence of paramagnetic-like behavior at the region surrounding the highlighted corners in Fig.~\ref{fig:sq-grid}. This re-entrance of the paramagnet is predicted by cluster mean-field (CMF) results~\cite{jin2016cluster-meanfield} and differs from the paramagnetic ordering along the $J_x = J_y$ line. For $J_x = J_y$, the anisotropic XYZ Heisenberg Hamiltonian reduces to an XXZ model that conserves total $\sigma^z$, leaving the coherent part of the dynamics unable to counteract the spin-lowering dissipation, and ultimately driving the system into a trivial all-down steady state. The re-entrance of the paramagnetic phase at large couplings is driven by a much different mechanism: As the interactions grow stronger, the state becomes increasingly mixed, eventually losing all transverse order and reverting to a paramagnetic regime.

\begin{figure}[h]
    \centering
    \includegraphics[scale=0.5]{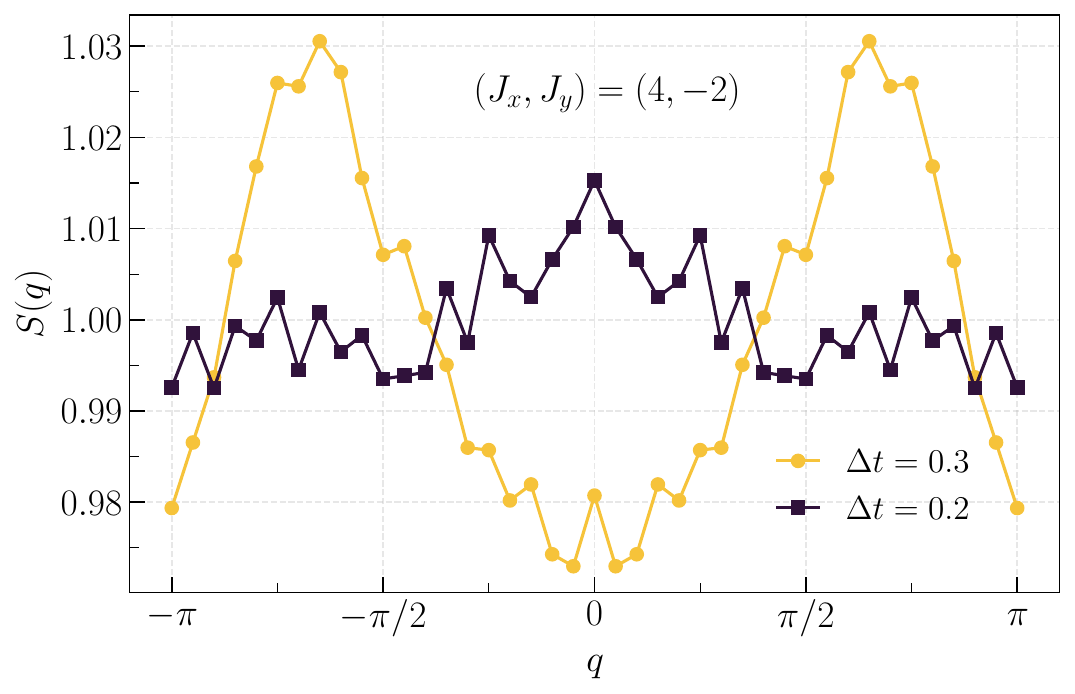}
    \caption{Static structure factor $S(q)$ at the unstable re-entrant paramagnetic phase for two different values of time step $\Delta t$. Simulated system is a periodic chain with $L=40$ sites, with coupling amplitudes $(J_x,J_y)=(4,-2)$ and dissipative rate $\gamma=J_z=1$.}
    \label{fig:dt_comparison}
\end{figure}

As predicted by cluster mean-field theory, this re-entrant paramagnetic regime is susceptible to finite-momentum instabilities, making it vulnerable to incommensurate orderings. The stability analysis of Ref.~\cite{jin2016cluster-meanfield} shows that once short-range correlations are included, the paramagnetic steady state can develop an unstable mode at nonzero momenta $q$, indicating a susceptibility toward incommensurate behavior. Our hardware simulations exhibit clear signatures of this effect at several points in the $(J_x,J_y)$ grid, as highlighted in the corners of Fig.~\ref{fig:sq-grid}. Interestingly, this appearance of incommensurate peaks is highly sensitive to the Trotter step size $\Delta t$. Because the Trotterization modifies the effective Liouvillian, small changes in $\Delta t$ can act as a perturbation that either preserves the paramagnetic character or pushes the system across the finite-momentum instability boundary (see also App.~\ref{app:incipient-sdw}). As can be seen in Fig.~\ref{fig:dt_comparison}, at $\Delta t = 0.2$ the system remains largely paramagnetic, with only a weak residual ferromagnetic component. Increasing the step size to $\Delta t = 0.3$ alters the effective dynamics enough to trigger the finite-momentum instability predicted by CMF theory, resulting in the emergence of clear incommensurate peaks. This behavior is consistent with the picture that the re-entrant paramagnet lies close to an instability boundary, and modest perturbations to the Liouvillian can determine whether the system remains in the mixed paramagnetic state or develops short-range incommensurate structure.

Taken together, the hardware results show broad quantitative agreement with the mean-field phase structure while simultaneously reflecting the key corrections introduced by cluster mean-field analysis. Along the $J_x = J_y$ line, we recover the expected paramagnetic behavior, but the ``widening'' of this region to nearby points predicted by mean-field does not appear. The SDW regions exhibit the correct directional structure, though only one of the two mean-field SDW wedges is recovered in each corner. Along the $J_y = 4$ and $-4$ lines, the hardware also shows faint SDW structure on the opposite side of the paramagnetic line, however these features are much less pronounced than the primary SDW wedges and do not form a full second wedge. Ferromagnetic and anti-ferromagnetic features are present as well, but their spatial extent is significantly reduced. The emergence of re‑entrant paramagnetic behavior at large couplings, with the appearance of finite‑momentum instabilities and the sensitivity of incommensurate peaks to $\Delta t$, further underscores that the hardware is probing the same dynamical mechanisms responsible for the modified topology of the CMF phase diagram~\cite{jin2016cluster-meanfield}. Overall, the hardware simulations reproduce the qualitative landscape of the dissipative Heisenberg model while exhibiting the distortions, suppressions, and instabilities expected once short‑range correlations and finite‑size effects are properly taken into account.

\section{Exact simulations for small system sizes}\label{app:exact-sims-L12}
\setcounter{figure}{0}

We discuss exact diagonalization results for various observables using the Liouvillian dynamics of a periodic spin chain with $L=12$ sites. The results presented in this section are used to compare the hardware results in the main text and tensor network formulation results that are discussed in the following section.

\begin{figure}[htpb]
    \includegraphics[width=0.85\textwidth]{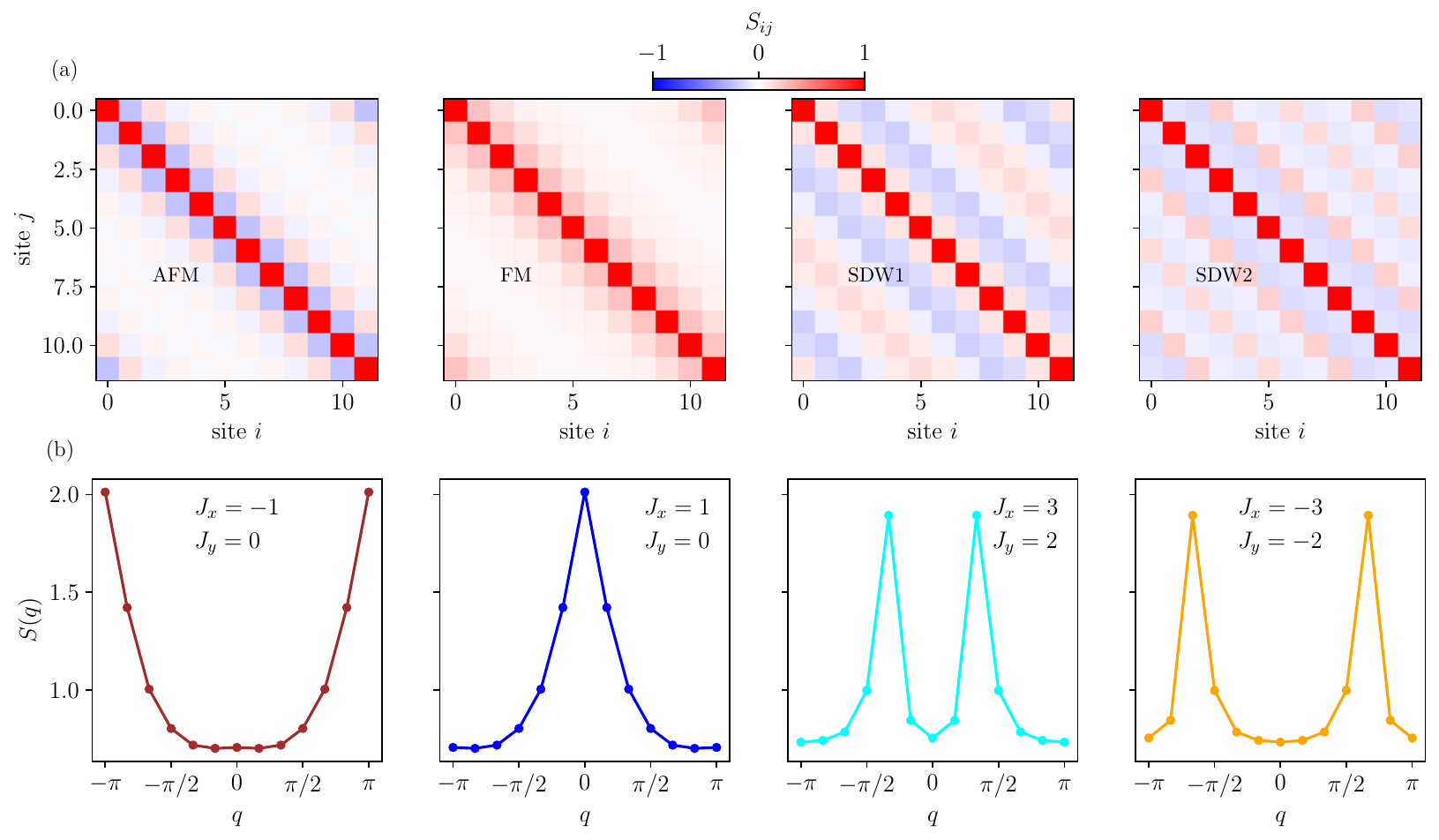}
    \caption{Steady state phases of the dissipative Heisenberg model obtained using exact diagonalization. Real space (a) and momentum space (b) structure factors for a 12 site chain.}
    \label{fig:exact_corr}
\end{figure}

\begin{figure}[htpb]
    \centering
    
    \includegraphics[width=0.55\textwidth]{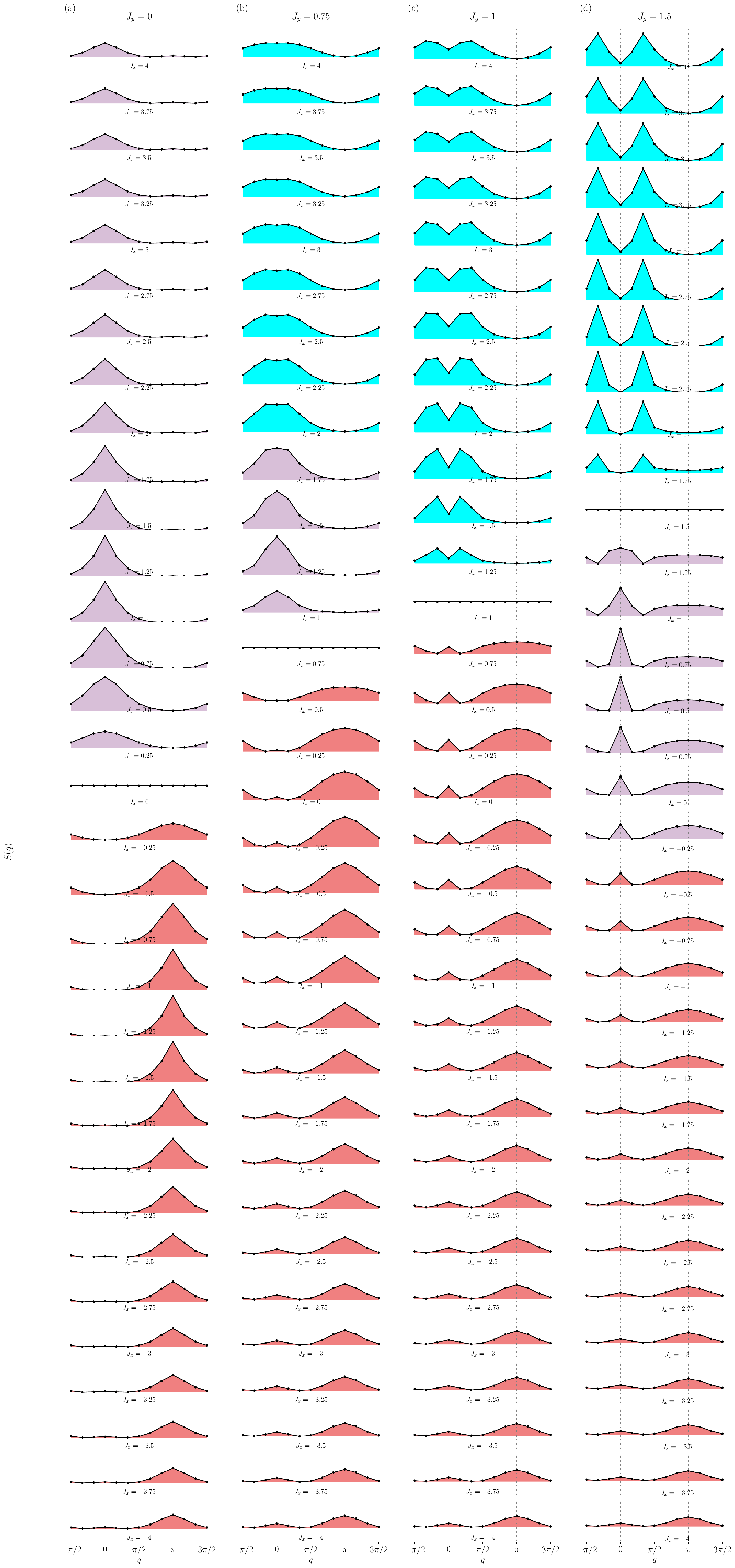}
    
    \caption{Observing the crossover between magnetic phases using exact diagonalization. Static structure factor $S(q)$ for several values of coupling amplitude $J_x$ and fixed (a) $J_y=0.0$, (b) $J_y=0.75$, (c) $J_y=1.0$ and (d) $J_y=1.5$. 
    \label{fig:exact_sq_Jx_scan}
    }
\end{figure}

\begin{figure}[!h]
    \centering
\includegraphics[width=0.9\textwidth]{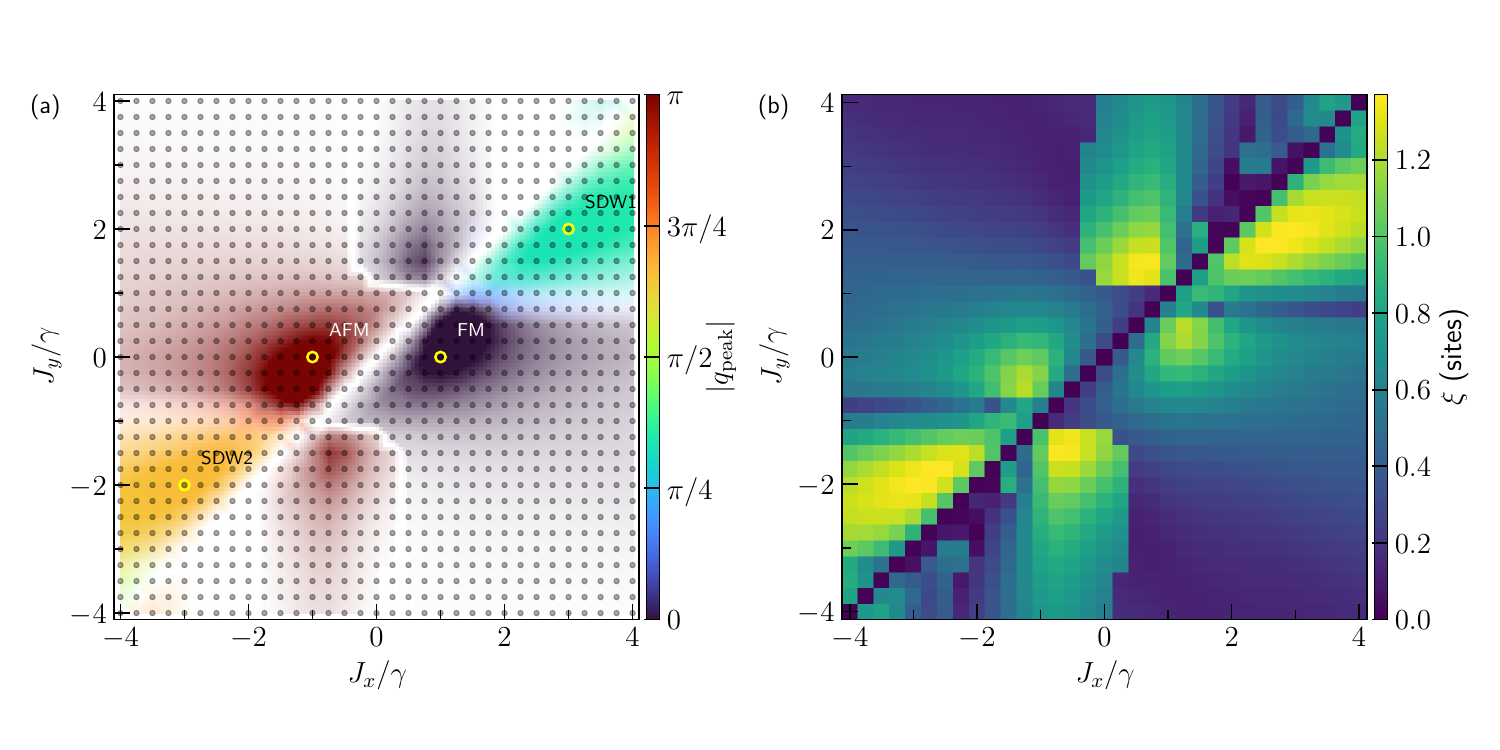}
    \caption{Phase diagram of the dissipative spin-1/2 Heisenberg model from the exact diagonalization calculation using (a) the peak position in structure factor $S(q)$, opacity is set by the height of the peak in the phasor construction described in the Appendix~\ref{app:interpolation} and (b) using the correlation length computed from the width of the structure factor. Data is interpolated before plotting. Since we use small Trotter step with $dt=0.05$, an incipient SDW phase is absent unlike the hardware results.
    \label{fig:exact_phase_diagram}
    }
\end{figure}

From the exact diagonalization calculation we can identify various magnetic phases. In Fig.~\ref{fig:exact_corr}, the real-space and momentum-space correlators of $\langle \sigma^x_i \sigma^x_j \rangle$ are plotted. Alternating blue and red bands identify the antiferromagnetic steady state, and a uniform red band identifies the ferromagnetic state. Both the real-space and momentum-space plots distinguish the spin density wave (SDW) from the other two phases: a periodicity greater than two in $S_{ij}$ identifies the SDW, and peaks away from $0$ and $\pi$ in momentum space identify the same phase. Interestingly, even for this small lattice the SDW1 and SDW2 phases show distinct differences, with the peaks shifted toward opposite sides of $q=\pi/2$. 
The peak height of $S(q)$ in the various phases identifies the strength of the different magnetic orderings, and this identification carries over to the structure factor plots in Fig.~\ref{fig:exact_sq_Jx_scan} as well. However, for the hardware data, the peak height is diminished and close to the case with the uniform average value of one, which is why distinguishing various phases near crossover regime becomes increasingly difficult as we go close to the boundary of various phases.  The figure shows that a scan at fixed $J_y$ not only recovers the various dissipative magnetically ordered phases but also indicates that several quantities associated with the structure factor have the potential to act as order parameters for identifying phase boundaries. The bulk phase identifications carry over to the large-scale hardware results.

Fig.~\ref{fig:exact_phase_diagram} shows various phases across the phase diagram, with white lines marking the phase boundaries or crossover regimes. The AFM, FM, SDW1, and SDW2 regions provide a rough idea of what to expect from larger-lattice simulations on the quantum simulator. The phase plots obtained from the peak prominence (a) and the correlation length (b) both identifies the phase boundaries and conform to the phase diagrams obtained from the quantum simulator. Phasor plot from the prominence can identify various phases, whereas phase diagram from the correlation length serves better to describe how strongly the system is correlated at various phases. The main difference from the quantum processing unit's result is the absence of the iSDW phases at the smaller Trotter step. The emergence of this phase in the hardware is explained in the Appendix~\ref{app:incipient-sdw}.

\section{Tensor network simulations}\label{app:tn-sims-L40}
\setcounter{figure}{0}

In this section we present our tensor network results for system size $L=40$.
Our classical simulations use the tensor jump method (TJM)~\cite{sander2025stochastic-simulation}, a tensor network adaptation of the Monte Carlo wavefunction method (MCWF)~\cite{molmer1993monte}. This technique that is a part of a family of methods called stochastic unraveling. The main idea rests on exchanging the deterministic Lindbladian evolution of the density matrix $\rho(t) \in \mathbb{C}^{2^L}\times \mathbb{C}^{2^L}$ with a stochastic Schrodinger evolution of a wavefunction $|\psi_i(t)\rangle \in \mathbb{C}^{2^L}$. Stochastic in nature, each evolution generates a different trajectory, and $\rho(t)$ is recovered by averaging over $|\psi_i(t)\rangle \langle \psi_i(t)|$. Each trajectory starts from an initial state that is subjected to a non-unitary evolution generated by a non-Hermitian Hamiltonian:
\begin{align}
    \tilde{H} = H + H_D,
\end{align}
where $H$ is the Hermitian system Hamiltonian, and $H_D$ is the dissipative Hamiltonian built from the jump operators:
\begin{align}
    H_D = -\frac{i\gamma}{2}\sum_m L_m^\dagger L_m.
\end{align}
We use this Hamiltonian to generate the Trotterized evolution
\begin{align}
    |\psi(t + \Delta t) ^{(i)} \rangle = e^{-i\tilde{H} \Delta t} |\psi_i (t)\rangle.
\end{align}
Because of the non-unitary of the evolution, the norm of the wavefunction decreases by some amount:
\begin{align}
    \langle \psi^{(i)}(t+\Delta t) |\psi^{(i)}(t+\Delta t) \rangle = 1- \delta p(t).
\end{align}
This denormalization is the direct result of the dissipative nature of the system, and in the picture of stochastic trajectories, is a sudden jump caused by the jump operators. The denormalization factor can be thought of as the combined effect of individual stochastic factors corresponding to our jump operators:
\begin{align}
    \delta p(t) = \sum_m \delta p_m(t),\quad \delta p_m(t) = \Delta t \gamma \langle \psi_i(t) |L^\dagger_m L_m| \psi_i(t) \rangle.
\end{align}
In this picture, $\delta p$ becomes the probability that a jump occurs at each time step. If no jump occurs, the wavefunction is normalized and we go to the next step:
\begin{align}
    |\psi_i(t+\Delta t)\rangle = \frac{1}{\sqrt{1-\delta p(t)}} |\psi^{(i)}(t+\Delta t)\rangle.
\end{align}
If a jump did occur, then we sample which particular process happened from the our set of jump operators, with each process having the probability $\delta p_m(t) / \delta p(t)$:
\begin{align}
    |\psi_i(t+\Delta t)\rangle = \sqrt{\frac{\gamma \Delta t}{\delta p_m(t)}}\ L_m |\psi_i(t)\rangle.
\end{align}
We then apply this process at every time step until the desired final simulation time is reached. This constitutes one trajectory.

TJM employs a tensor network scheme combined with a dynamic time-dependent variational principle (TDVP) algorithm to efficiently implement the MCWF method. One trajectory of TJM follows the evolution
\begin{align}
    U(T) = \prod_{i=0}^n \mathcal{F}_{n-i}(\Delta t),
\end{align}
where
\begin{align}
    \mathcal{F}_j(\Delta t) =
    \begin{cases}
        \mathcal{J}_\epsilon(\Delta t)\mathcal{D}(\Delta t/2) \mathcal{U}(\Delta t),\quad &j=n \\
        \mathcal{J}_\epsilon(\Delta t)\mathcal{D}(\Delta t) \mathcal{U}(\Delta t),\quad &0<j<n \\
        \mathcal{J}_\epsilon(\Delta t)\mathcal{D}(\Delta t / 2),\quad &j=0.
    \end{cases}
\end{align}
The unitary part of the evolution is encoded by $\mathcal{U}(\Delta t) = e^{-iH_0 \Delta t}$, and the dissipative evolution is encoded by $\mathcal{D}(\Delta t) = e^{-iH_D \Delta t}$, while the stochastic jump process is encoded by $\mathcal{J}_\epsilon(\Delta t)$.

The states are encoded as matrix product states (MPS) and the operators are encoded as matrix product operators (MPO). The unitary part of the evolution is computed using a dynamic TDVP method, where the solver dynamically switches between one-site TDVP and two-site TDVP based on whether the bond dimension has more room to grow. This allows for more smaller truncation errors while maintaining computational efficiency. The dissipative evolution and the stochastic jump evolution can also be efficiently encoded for single-site jump operators.

To cut down on memory requirements, we can save trajectories of expectation values instead of trajectories of full state vectors. If we are interested in a set of operators $\{O_j (t)\}$, then we run $N$ TJM trajectories, and we estimate the expectation values by an averaging process over the trajectories:
\begin{align}
    \langle O_j(t)\rangle = \frac{1}{N}\sum_{i=1}^N \langle\psi_i(t) |O_j | \psi_i(t) \rangle.
\end{align}
Since each TJM trajectory is an independent run, this algorithm is embarrassingly parallelizable.

\begin{figure}
    \includegraphics[width=0.9\textwidth]{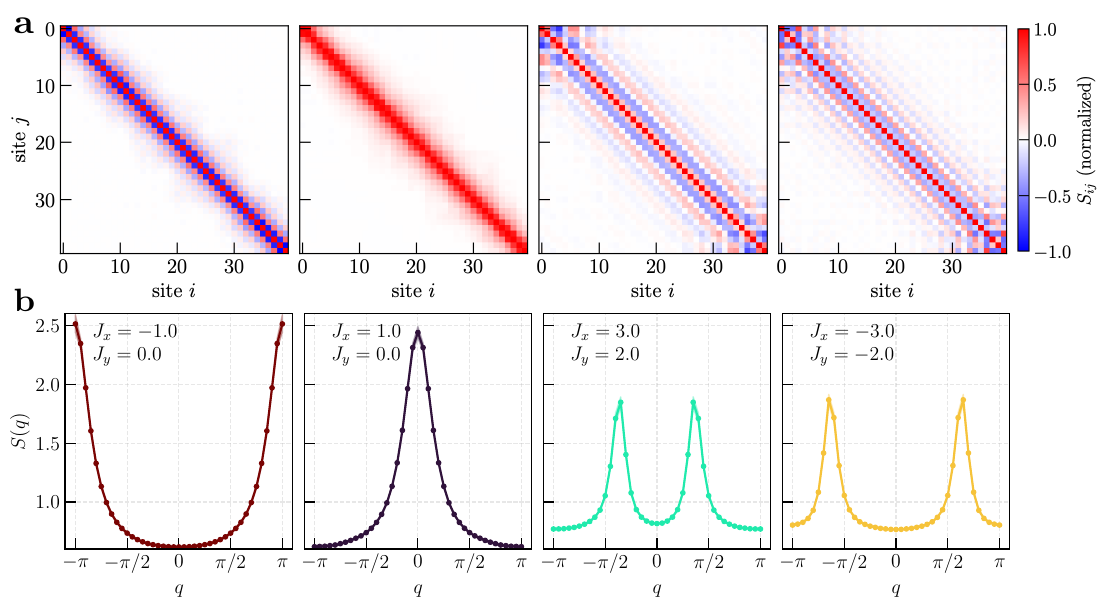}
    \caption{Real space spin-spin correlation function (a) and momentum space structure factor (b) for a 40-site chain obtained using the tensor jump method described in the text. Shaded area around the curves correspond to the 95\% confidence interval of 1000 bootstrap samples, resampled from the 128 trajectories.}
    \label{fig:tjm}
\end{figure}

For the simulations in this paper, we used the python package provided by the authors of TJM~\cite{mqt}. The original code accepted only 1-site and neighboring 2-site operators, so we modified the code to allow for arbitrary 2-site operators to be able to determine the structure factors~\cite{yaqs-modified}. We ran the simulations for $L=40$ with final simulation time $T=20$ and time step $\Delta t=0.1$. The bond dimension was capped at 16, and the averages were calculated over 128 trajectories. Correlator and structure factor plots of the NESS is shown in  the Fig~\ref{fig:tjm} which shows qualitatively similar behavior to exact results at smaller lattice. The location of the peaks in the structure factor plots are in agreement with hardware results for the parameters we ran, but the amplitudes from the hardware data appear to be damped. Each parameter set took approximately 25 minutes on an Apple M3 Pro chip using 10 threads. We also ran the same parameters for bond dimension caps 8 and 32 (not shown in the paper), and the results remained largely consistent with what was shown. The locations of the peaks were the same, but there was a very slight change in the amplitudes of the structure factor points.

\section{Interpolation of the phase diagram}\label{app:interpolation}
\setcounter{figure}{0}

In this section we provide details about the interpolation scheme employed in sketching the phase diagram based on the peak position and height of the static structure factor in momentum space.
For every measured coupling pair $(J_x, J_y)$ we build the equal-time $S_{ij} = \mathrm{Tr} \left[ \rho \sigma^x_i \sigma^x_j \right]$
correlation matrix averaged over the last ten Trotter steps, and Fourier transform to the static structure factor $S(q)$ with
$q \in [0, \pi]$:
\begin{align}
    S(q) = \frac{1}{L}\sum_{i,j} S_{ij} e^{iq(r_i-r_j)}
\end{align}

Each point is then reduced to two descriptors: the peak
wavevector $q^{\ast}$, the location of the dominant maximum of $S(q)$ refined
below the FFT bin spacing by a three-point parabolic fit, and its prominence
$\Pi = S(q^{\ast}) - \operatorname{median}_q S(q)$.  Hardware data carry a
common-mode artifact confined to the $q = 0$ bin; when $S(0)$ forms a
single-bin cusp that overshoots the smooth trend extrapolated from its
neighbors, that bin is excluded from the peak search so that $q^{\ast}$
tracks the physical ordering peak.

The field $q^{\ast}(J_x, J_y)$ cannot be interpolated as a scalar: across a
direct FM--AFM boundary it jumps from $0$ to $\pi$ with no intermediate
ordering in between, and linear interpolation would sweep through every
intermediate wavevector, painting a spurious SDW-colored band along a
boundary where the data show no such order.  Instead, each measured point is
mapped to a prominence-weighted phasor
\begin{equation}
  \mathbf{v} = \Pi \,\bigl( \cos q^{\ast},\, \sin q^{\ast} \bigr),
  \label{eq:phasor}
\end{equation}
and the two components are interpolated linearly onto the plot grid.  At each
grid cell the color encodes the recovered angle
$q = \operatorname{atan2}(v_y, v_x)$, which remains in $[0, \pi]$ because
$v_y \ge 0$, and the opacity is set by $|\mathbf{v}|$.  Where neighboring
points order at similar wavevectors this reduces to ordinary linear
interpolation of $q^{\ast}$ and $\Pi$, so smooth regions --- including the
genuine incommensurate lobes --- are rendered exactly as a naive scheme
would.  Across an FM--AFM boundary, however, the two phasors are
antiparallel: the interpolated $\mathbf{v}$ moves along the axis joining
them, the recovered angle switches directly from $0$ to $\pi$ without
visiting intermediate values, and $|\mathbf{v}|$ dips to zero.  Such boundaries therefore appear as thin transparent seams; the seam
marks a switch in the identity of the dominant peak, not an absence of order.

The opacity ramps linearly from zero at a prominence floor $\Pi_0 = 0.03$,
below which we consider no order discernible, to one at the 95th percentile
of the measured prominences.  Outside the convex hull of the measured points
the map is fully transparent.  As a cross-check of the scheme, we also
interpolated the baseline-subtracted $S(q)$ profiles themselves, bin by bin
in $q$, and re-extracted the dominant peak at every grid cell; away from
phase boundaries the two constructions agree, and both eliminate the phantom
boundary colors produced by scalar interpolation of $q^{\ast}$.

In the far anti-diagonal corners a weak but reproducible incommensurate
signal (symmetry partners under $(J_x, J_y) \to (-J_x, -J_y)$) is nearly
invisible on the global opacity scale.  Inside the corner triangles
$|J_x - J_y| \ge 5.5$ the opacity is therefore stretched onto a local scale,
saturating at the 95th percentile of the corner-point prominences with a
ramp four times steeper than the global one; the dashed lines in the figure
mark this change of scale.

\section{Higher-order Trotter terms and the incipient SDW phases}\label{app:incipient-sdw}
\setcounter{figure}{0}

In this section we discuss how by setting a relatively large Trotter step we observe incipient SDW phases in the main diagonal corners of the phase diagram, which arise as a consequence of higher-order corrections to the Trotter approximation.

One circuit step applies the unitary even bond layer $\mathcal{L}_e$, the unitary odd bond layer $\mathcal{L}_o$, and the
dissipation $\mathcal{L}_D$ in sequence, $e^{\mathcal{L}\,\Delta t}\simeq
e^{\mathcal{L}_e\,\Delta t}\,e^{\mathcal{L}_o\,\Delta t}\,
e^{\mathcal{L}_D\,\Delta t}$, so the leading correction to the generator is
\begin{align}
  \mathcal{L}_{\rm eff}
  = \mathcal{L}_H + \mathcal{L}_D
  + \frac{\Delta t}{2}\Bigl(\bigl[\mathcal{L}_e,\mathcal{L}_o\bigr]
  + \bigl[\mathcal{L}_H,\mathcal{L}_D\bigr]\Bigr)
  + \mathcal{O}(\Delta t^{2}),
  \qquad \mathcal{L}_H = \mathcal{L}_e + \mathcal{L}_o .
  \label{eq:trotter-generator}
\end{align}
The first commutator is the Hamiltonian correction $\Delta H$. Splitting the Hamiltonian into even and odd bond layers,
$e^{-iH\,\Delta t}\simeq e^{-iH_e\,\Delta t}\,e^{-iH_o\,\Delta t}$, the
circuit implements the effective Hamiltonian
\begin{align}
  H_{\rm eff} = H + \Delta H + \mathcal{O}(\Delta t^{2}),
  \qquad
  \Delta H = -\frac{i\,\Delta t}{2}\,[H_e, H_o].
  \label{eq:trotter-splitting}
\end{align}
Only adjacent bonds sharing a site contribute to the commutator.  With
$[\sigma^a_m,\sigma^b_m] = 2i\,\epsilon_{abc}\,\sigma^c_m$ on the shared site,
each pair of couplings $J_a J_b$ ($a\neq b$) generates a three-site term, and
collecting all of them gives
\begin{align}
  \Delta H = \Delta t \sum_{m} (-1)^{m+1} \Bigl[\,
      & J_x J_y \bigl( X_{m-1} Z_m Y_{m+1} - Y_{m-1} Z_m X_{m+1} \bigr) \\
    + & J_y J_z \bigl( Y_{m-1} X_m Z_{m+1} - Z_{m-1} X_m Y_{m+1} \bigr) \\
    + & J_z J_x \bigl( Z_{m-1} Y_m X_{m+1} - X_{m-1} Y_m Z_{m+1} \bigr)
  \,\Bigr].
  \label{eq:delta-H}
\end{align}
All three families are chiral: three-site, staggered, and odd under
reflection about the central site.  At the isotropic point
$J_x = J_y = J_z = J$ the bracket in Eq.~\eqref{eq:delta-H} collapses to
$-8J^{2}\,\mathbf{S}_{m-1}\cdot(\mathbf{S}_{m}\times\mathbf{S}_{m+1})$, the
staggered scalar spin chirality; away from it the three families carry
independent weights $J_aJ_b$.  Along the $J_y = 0$ line the $xy$ and $yz$
families vanish identically and only the $zx$ family survives.

The second commutator in Eq.~\ref{eq:trotter-generator}
is not a Hamiltonian term: for the jump operators $\sigma^-_i$ at rate $\gamma$,
\begin{align}
  \bigl[\mathcal{L}_H,\mathcal{L}_D\bigr](\rho)
  = \gamma \sum_i \Bigl[
      \underbrace{-i\Bigl( [H,\sigma^-_i]\,\rho\,\sigma^+_i
                         + \sigma^-_i\,\rho\,[H,\sigma^+_i] \Bigr)}_{\text{modified jump}}
      \;+\;
      \underbrace{\frac{i}{2}\bigl\{\,[H,\,\sigma^+_i\sigma^-_i],\,\rho\,\bigr\}}_{\text{modified damping}}
    \Bigr].
  \label{eq:cross-term}
\end{align}
With $\sigma^-_i = (X_i - iY_i)/2$ and $\sigma^+_i\sigma^-_i = (1+Z_i)/2$ the
required commutators are two-site,
\begin{align}
  [H,\sigma^-_i]
  = \sum_{j\,\in\,{\rm nn}(i)}
    \Bigl( J_x\, Z_i X_j \;-\; i J_y\, Z_i Y_j \;-\; 2 J_z\, \sigma^-_i Z_j \Bigr),
  \qquad
  [H,\,\sigma^+_i\sigma^-_i] = \tfrac{1}{2}\,[H, Z_i],
  \label{eq:H-sigma-commutator}
\end{align}
so the Trotterized dissipation is no longer local: the step generates
correlated two-site decay together with a state-dependent damping
$\propto [H, Z_i]$.

Equivalently, to first order in $\Delta t$ the entire cross term amounts to
keeping the Lindblad form and renormalizing each jump operator,
\begin{align}
  \tilde{\sigma}^-_i
  = \sigma^-_i - \frac{i\,\Delta t}{2}\,[H,\sigma^-_i]
  = \sigma^-_i - \frac{i\,\Delta t}{2}
    \sum_{j\,\in\,{\rm nn}(i)}
    \Bigl( J_x\, Z_i X_j - i J_y\, Z_i Y_j - 2 J_z\, \sigma^-_i Z_j \Bigr),
  \label{eq:modified-jump}
\end{align}
i.e.\ $\mathcal{D}[\sigma^-_i] \to \mathcal{D}[\tilde{\sigma}^-_i]$ reproduces
Eq.~\eqref{eq:cross-term} up to $\mathcal{O}(\Delta t^{2})$.

\begin{figure}[h]
    \includegraphics[clip=true,trim=0 0 0 50,width=0.95\textwidth]{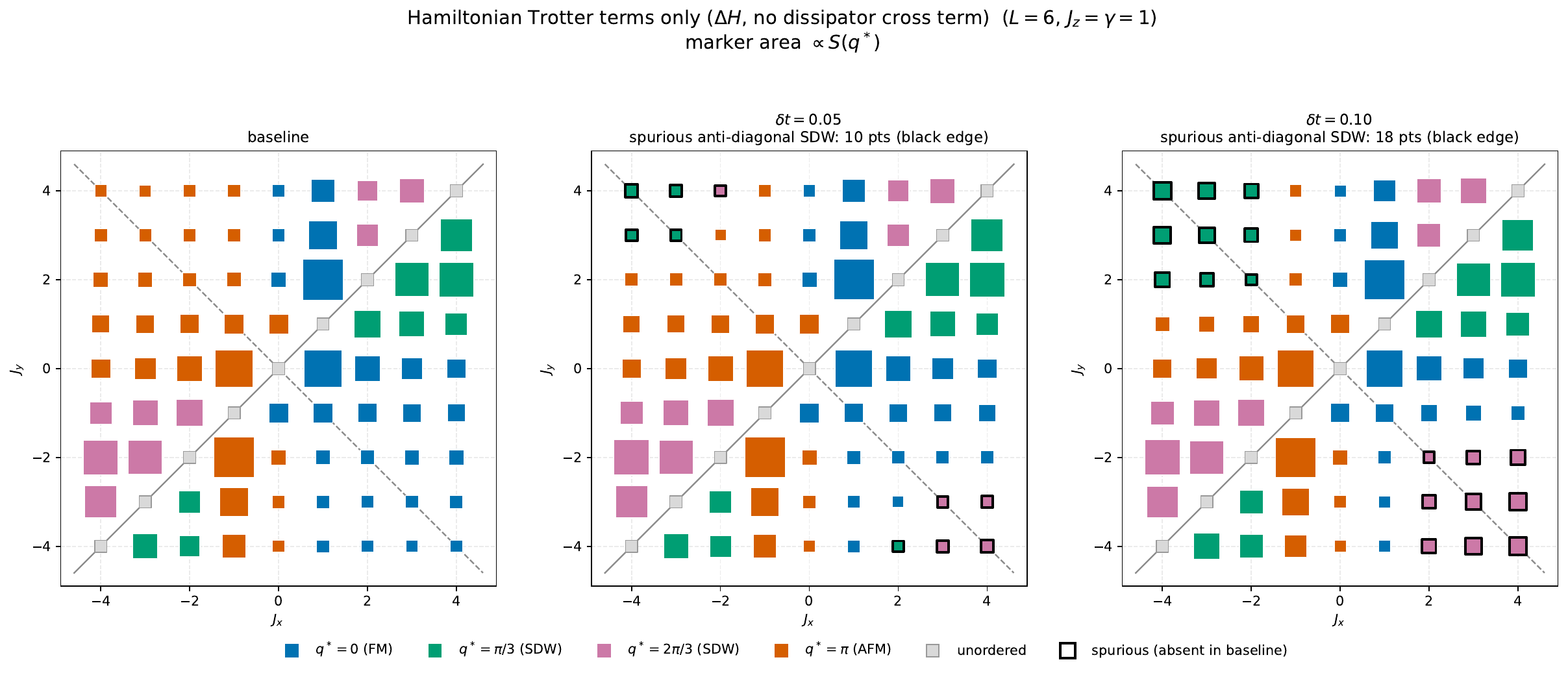}
    \caption{Location of the peak in the spin structure factor for an $L=6$ dissipative Heisenberg chain, for baseline (no Trotter error), and $\Delta t=0.05$ and $\Delta t=0.10$. The size of the markers indicates the relative strength of the order (peak height), whereas the color indicates the peak position.}
    \label{fig:trotter-dt-series-dH}
\end{figure}

In Fig.~\ref{fig:trotter-dt-series-dH} we show that the additional terms in $\Delta H$ give rise to the incipient SDW phases in the corners of the phase diagram. We use small scale ($L=6$) exact diagonalization of the Liouvillian to determine the steady state. As the Trotter error is increased from $0$, the iSDW phases emerge in the corners, and this effect gets stronger with increasing $\Delta t$.  The rest of the phase diagram remains unaffected by the Trotter error, in agreement with the quantum hardware results.
The modification of the jump operator $\sigma^-_i \rightarrow \tilde \sigma^-_i$, while it affects the strength of the observed orders, does not shift the peak.

\end{document}